\documentclass[aps,prd,twocolumn,groupedaddress,nofootinbib]{revtex4-1}
\pdfoutput=1  
\usepackage{bm}
\usepackage{graphicx,color}
\usepackage{latexsym,amsmath,amssymb,graphicx,booktabs}
\usepackage{hyperref}
\usepackage{placeins}
\usepackage{subfigure}
\usepackage{scrextend}

\definecolor{MyBlue}{rgb}{0.15,0.15,0.70}
\definecolor{Dgreen}{rgb}{0,0.7,0.0}

\hypersetup{
colorlinks=true,
citecolor=MyBlue,
linkcolor=MyBlue,
urlcolor=MyBlue
}

\usepackage{amssymb}
\usepackage{amsmath}
\usepackage{amsfonts}
\usepackage{upgreek}
\usepackage{latexsym}
\usepackage{appendix}
\usepackage{eufrak}
\usepackage{dsfont}

\usepackage[export]{adjustbox}

\include{mydefs}

\newcommand\spart{\;\raise1.0pt\hbox{/}\hskip-6pt\partial}
\newcommand\spartb{\;\overline{\raise1.0pt\hbox{/}\hskip-6pt\partial}}

\newcommand{\be}{\begin{equation}}
\newcommand{\ee}{\end{equation}}
\newcommand{\bk}{{\mathbf k}}

\newcommand{\beqa}{\begin{eqnarray}}
\newcommand{\eeqa}{\end{eqnarray}}

\newcommand{\bx}{{\bf{x}}}
\newcommand{\bw}{{\bf{w}}}
\newcommand{\bv}{{\bf{v}}}
\newcommand{\bW}{{\bf{W}}}

\newcommand{\bu}{{\bf{u}}}
\newcommand{\bpsi}{{\bf{\Psi}}}
\newcommand{\bq}{{\bf{q}}}
\newcommand{\bp}{{\bf{p}}}
\newcommand{\bJ}{{\bf{J}}}

\newcommand{\bGamma}{{\bf{\Gamma}}}

\renewcommand{\AA}{\mathcal{A}}

\newcommand{\HH}{\mathcal{H}}
\newcommand{\ep}{\epsilon}
\newcommand{\de}{\delta}
\newcommand{\De}{\Delta}
\newcommand{\si}{\sigma}
\newcommand{\we}{\wedge}
\newcommand{\opt}{{\hat{ \cal T}}}
\newcommand{\al}{\alpha}
\renewcommand{\b}{\beta}
\newcommand{\ga}{\gamma}
\newcommand{\Ga}{\Gamma}
\newcommand{\Om}{\Omega}
\newcommand{\La}{\Lambda}
\newcommand*{\tin}{_{\text{in}}}
\newcommand*{\ti}{\tilde}
\newcommand{\trr}{\text{Tr}}
\newcommand{\dd}{\text{d}}
\newcommand{\pd}{\partial}
\newcommand{\na}{\nabla}
\newcommand{\nn}{\nonumber}
\newcommand{\bss}{\boldsymbol{\mathcal{S}}}
\newcommand{\bom}{\boldsymbol{\omega}}

\newcommand{\ra}{\rightarrow}
\newcommand{\xra}[1]{\xrightarrow[#1]{}}

\graphicspath{ {./Graphs/} }
\begin{document}

\vspace*{2cm}

\title{Vorticity  generation  in the Universe: A perturbative approach}
\author{Giulia Cusin, Vittorio Tansella and Ruth Durrer}
\affiliation{D\'epartement de Physique Th\'eorique and Center for Astroparticle Physics, Universit\'e de Gen\`eve, 24 quai Ansermet, CH--1211 Gen\`eve 4, Switzerland}
\email{giulia.cusin@unige.ch, vittorio.tansella@unige.ch, ruth.durrer@unige.ch}
\vspace{1 em}
\date{\today}

\begin{abstract}
We  compute the generation of vorticity from velocity dispersion in the dark matter fluid. For dark matter at zero temperature Helmholtz's theorem dictates that no vorticity is generated and we therefore allow the dark matter fluid to have a non-vanishing velocity dispersion. This implies a modification to the usual hydrodynamical system (continuity \& Euler equations): we have to consider the Boltzmann hierarchy up to the second moment. This means that the Euler equation is modified with a source term that describes the effect of non-zero velocity dispersion. We write an equation for the Eulerian vorticity in Lagrangian coordinates and show that it has a growing mode already at second order in perturbation theory. 
We compute the power spectrum of the vorticity and the rotational velocity at second order in perturbation theory.
\end{abstract}

\maketitle


\section{Introduction}
It is well known that in the observed Universe galaxies rotate. Also clusters of galaxies have a non-vanishing vorticity and it has recently been argued that vorticity is correlated on even  larger scales,  up to 20$h^{-1}$Mpc~\cite{Taylor:2016rsd}. It is well known from numerical simulations that strong nonlinearities in the dark matter (DM) evolution lead to `shell crossing' which can induce vorticity.  It is, however, not evident that vorticity with such a large correlation scale can be generated by  shell crossing.  

Standard cosmological perturbation theory of pressureless matter does not predict generation of vorticity. Even though the momentum (at second order) is not curl free, $\nabla\wedge(\rho\bv)\neq 0$, the velocity field is, $\nabla\wedge\bv=0$. This is a simple consequence of Helmholtz's theorem on vorticity conservation in a perfect fluid: in the absence of rotational external forces (as is the case in newtonian gravity), a fluid that is initially irrotational remains irrotational\footnote{Vorticity conservation generalizes to general relativity where vorticity is  the antisymmetric part of the covariant derivative of the four velocity of matter~\cite{Hollenstein:2007kg,Lu:2008ju}. Even though, the gravitational field (i.e. the space-time metric) in general also contains a vector part, leading to the relativistic effect of frame dragging, this comes from the vector component of the fluid momentum density and not from vorticity~\cite{Lu:2008ju,Adamek:2015eda}. See also \cite{Rampf:2016wom,Adamek:2016zes} for a detailed discussion.}. In the present work, we focus on the Newtonian case. It has been explicitly verified that in the Lagrangian and the Eulerian approach, as long as particles are accelerated by Newtonian gravity alone, no vorticity is generated at any order, see e.g.~\cite{Bernardeau:2001qr}. On the other hand, Newtonian N-body simulations show that vorticity is generated, see e.g. \cite{Pueblas:2008uv}, \cite{Paduroiu:2015jfa}. It is not clear if this is due to shell crossing only or also to velocity dispersion which, even if not present in the initial conditions, does build up during the evolution of the system due to finite resolution.

When shell crossing occurs, the velocity in a given fluid element is no longer single valued: the standard Eulerian and Lagrangian descriptions break down and vorticity can be generated.  But can it be generated before in a fluid of free streaming particles moving under Newtonian gravity? And, more relevant for this work: can we understand the generation of vorticity perturbatively?

\begin{figure*}
\includegraphics[scale=0.4]{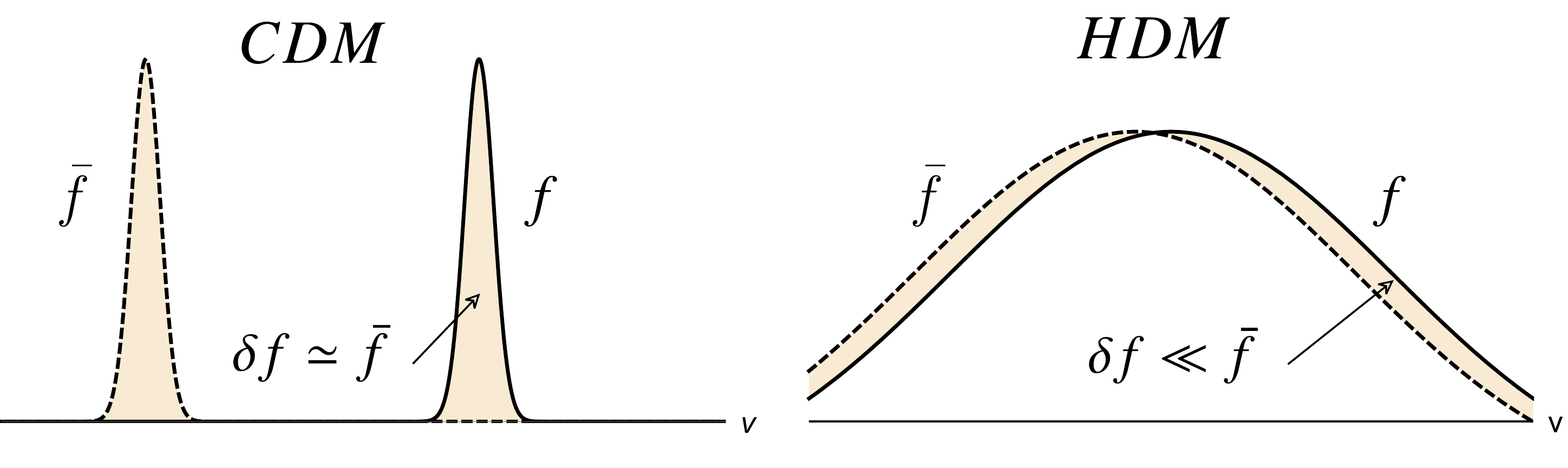}
\caption{\label{f:motion}
The center of the distribution function may move away from its origin under the action of gravitational acceleration, but its width changes very little. For cold dark matter (\emph{left}) this means that in the split $f= \bar f + \de f (\bx)$ the perturbation $\de f$ is similar in magnitude to $\bar f$. For hot species (\emph{right}) the width of the distribution allows to treat $\de f$ as a perturbation since $\de f \ll \bar f $.}
\end{figure*}

An important ingredient needed for the generation of vorticity is the presence of velocity dispersion. (Note that also shell crossing can be regarded as a kind of `velocity dispersion' since when it is present, the velocity takes on different values at certain positions.)  As long as the velocity at a given position has a deterministic value which is accelerated by the gradient of the Newtonian potential, vorticity is not generated \cite{Ehlers:1996wg}. However, if the fluid velocity is an average over a momentum distribution, this averaging procedure can lead to vorticity.  We want to address vorticity generation perturbatively. A first attempt might be the perturbed Vlasov equation. However, the dark matter flow is so cold that the distribution function $f$ in momentum space is nearly a Dirac-delta function so that even a small gravitational acceleration leads to large changes in the center of the 
distribution function in momentum space and in an ansatz $f=\bar f(|\bp|,t) + \de f(\bp,\bx,t)$,   
$\de f$ does not stay small, see Fig.~\ref{f:motion}. However, the width of $f$ in velocity space is very small and we can use it as our small quantity.  
This is the plan of the present work. We discuss vorticity generation from velocity dispersion perturbatively. We show that starting from purely scalar perturbations with vanishing vorticity, at second order in perturbation theory  vorticity is generated, with an amplitude  proportional to the $0$th order amplitude of the velocity dispersion.  In a matter dominated Universe, this vorticity  grows like the square root of the linear density growing mode and we compute its power spectrum. 

The remainder of the paper is structured as follows: in the next section we develop the Eulerian and Lagrangian schemes for Newtonian gravitational clustering in the presence of velocity dispersion, concentrating on the equation of motion of vorticity. Eulerian and Lagrangian perturbation theory are two different but equivalent approaches. While the former deals with fields at a fixed space-time position $\bx$, the latter follows particle trajectories by introducing a displacement field $\bss $ used in the mapping $\bx = \bq +\bss$ between the initial position $\bq$ and the Eulerian position $\bx$. We will derive the fundamental equations in Section~\ref{ss:VlaApp} using the Eulerian approach while in the rest of the paper we present their solutions in the Lagrangian scheme. To make the work more self contained, technical details and several derivations which can  also be found in the literature are presented in appendices. In Section~\ref{s:res} we present our results for a matter dominated Universe. In Section~\ref{s:con} we discuss our findings and conclude.\\

{\bf{Notation:}}  We work in a spatially flat Friedmann-Lema\^\i tre (FL) Universe with metric
\begin{align*}
ds^2 &=a^2(\eta)\left(-d\eta^2+\de_{ij}dx^idx^j\right) \\
&= -dt^2 +a^2(t)\de_{ij}dx^idx^j\,.
\end{align*}
Here $\eta$ is conformal time while $t$ is called `physical time'. We normalize the present scale factor to unity, $a_0=a(\eta_0)=a(t_0)=1$. Furthermore, we denote 3d vectors in bold face and  we use the equivalent notations $\na f= \na_\bq f$ or $\na_i A^i = A_{i,i} =\na \cdot \mathbf{A}$ for derivatives with respect to the Lagrangian position $\bq$, while we always specify the derivative with respect to Eulerian coordinates: $\na_\bx f$ or $\na_\bx \cdot \mathbf{A} $.

\section{Generation of vorticity by velocity dispersion}

\subsection{The Vlasov approach with velocity dispersion} \label{ss:VlaApp}

We consider collisionless massive particles moving under their own (Newtonian) gravity.
Newtonian gravity is a long range force and is well described by the Vlasov equation
\be\label{e:Vlasov}
\frac{\dd f}{\dd \eta}=\left(\frac{\partial f}{\partial \eta}\right)_{\bx}+\frac{\bp}{ma}\cdot \nabla_\bx f-m a \nabla_\bx \Phi\cdot \frac{\partial f}{\partial \bp}=0\,.
\ee
Here $f(\eta,\bx,\bp)$ is the (non-relativistic) particle distribution function in phase space, and the factors $a$ stem from the fact that $\bp \equiv m a^2 \, \dd \bx / \dd t =   m a \, \dd \bx / \dd \eta = m a \,  \bu$ is the redshift corrected momentum and $\bu$ is the peculiar velocity. $\Phi(\bx,\eta)$ is the gravitational potential satisfying the Poisson equation
\be\label{e:Poisson}
\De\Phi = 4\pi\de\rho \,,
\ee
where 
\be
\rho =\bar\rho +\de\rho = \bar\rho(1+\de)\,.
\ee
We also split $f(\eta,\bx,\bp) =\bar f(\eta,|\bp|) + \de f(\eta,\bx,\bp)$. The overdensity $\de$ is determined self-consistently from $\de f$, however we do not request that $|\de f|\leq \bar f$ pointwise in momentum space. As we have mentioned in the introduction, since the velocity dispersion is very small, even a modest gravitational force can move the peak of $f$ into a different region of velocity space and lead to large deviations from $\bar f$, see fig.~\ref{f:motion}. Nevertheless, we do expect the velocities to stay small (albeit much larger than the velocity dispersion) so that that the momentum integrals change little. In particular, we expect also the width of $f$ in velocity space to change little so that the integrals of $\de f$ over velocities remain small.

We can construct moments of the Vlasov equation by integrating~\eqref{e:Vlasov} weighted by products of $p^i$ over $\dd^3 p$ and by using the fact that boundary terms vanish since $f(\eta,\bx,\bp)\xra{|\bp|\ra\infty} 0$. For example the $n$th moment yields
\be
\int \! \! \dd^3 p \,\,\frac{df}{d\eta}(\eta,\bx,\bp) \,p^{i_1}\cdots p^{i_n} \, = \, 0\,.
\ee
For the zero-th moment we obtain the continuity equation,
\be\label{e:con}
\pd_\eta\de + \nabla_\bx((1+\de)\bv)=0 \,,
\ee 
where we have introduced
\beqa
\rho(\eta,\bx)\!\! &=&\!\! \frac{m}{a^3}\int \! \! \dd^3p\, f(\eta,\bx,\bp)\! =\bar\rho(\eta)[1\!+\de(\eta,\bx)] \\
\bar\rho(\eta) &=& \!\!\frac{m}{a^3}\int \!\!\dd^3p\, \bar f(p)\,,
\eeqa
and
\be
(1+\de)\bv(\eta,\bx) = \frac{1}{\bar\rho a^4}\int \dd^3p\, \bp f(\eta,\bx,\bp) \,.
\ee
The first moment of the Vlasov equation gives the Euler equation,
\begin{align}\label{e:Euler}
&\pd_\eta[(1+\de)v_i] +\HH(1+\de)v_i +\nn\\
&+\pd_j[(1+\de)(v_iv_j+\si_{ij})] +(1+\de)\pd_i\Phi=0 \,,
\end{align}
where we have introduced the velocity dispersion tensor (VDT)
\be
\si_{ij} = \langle u_iu_j\rangle_p -  \langle u_i\rangle_p  \langle u_j\rangle_p  \,.
\ee
Here the $ \langle \cdots \rangle_p $ indicates an average over momentum space, i.e. for an observable $\AA(\eta,\bx,\bp)$ in phase space we define
\be
\langle \AA\rangle_\text{p}(\eta,\bx) \equiv \frac{\int d^3p \, \AA(\eta,\bx,\bp) f(\eta, \bx, \bp)}{\int d^3p f(\eta, \bx, \bp)}\,,
\ee
so that $\langle u_i\rangle_p=v_i= (am)^{-1} \langle p_i \rangle$.
The third term in the Euler equation (\ref{e:Euler}) is simply the stress tensor,
\be
T_{ij} = P\de_{ij}+\Pi_{ij} = \rho(v_iv_j+\si_{ij}) \,,
\ee
where $P=\rho(v^2+\si_i^i)/3$ denotes the pressure and $\Pi_{ij}$ is the anisotropic stress tensor.\footnote{We observe that the trace part of the VDT tensor $\sigma_{ij}$ corresponds to pressure while the  off-diagonal part is the anisotropic stress of the dark matter fluid.} If the velocity dispersion vanishes we have a perfect fluid and, as it is well know (Helmhotz' theorem) no vorticity is generated. However, this is no longer the case if $\si_{ij}\neq 0$.
Contrary to $v_i$,  the tensor $\si_{ij}$ already has a zero-order contribution,
\be
\si^{(0)}_{ij} = \si^{(0)}\de_{ij}\,,
\ee
where 
\be
\si^{(0)} = \frac{1}{3a^5\bar\rho}\int d^3p\, \frac{p^2}{m} \bar f(\eta,\bp)\,.
\ee
We assume that this effective pressure from velocity dispersion is small so that we can neglect its contribution to the Friedmann equation governing the background evolution.
If $\bar f$ is a Maxwell Boltzmann distribution at temperature $T$ then $\si^{(0)} =T/m$ and $P=\rho\si^{(0)}$ is the fluid pressure.

Using (\ref{e:con}) to eliminate time derivatives of the density perturbation, we can rewrite the Euler eq.~(\ref{e:Euler}) in the form
\be\label{e:Euler2}
\left(\pd_\eta +v^i\pd_i\right)v_j+\HH v_j = \pd_j\Phi-\frac{1}{\rho}\pd_i(\rho\si_{ij})\,.
\ee
Integrating the second moment of the Vlasov equation we obtain the third member of the collisionless Boltzmann hierarchy. Using \eqref{e:con} and \eqref{e:Euler2} we can again eliminate the time derivatives of $\de$ and $\bv$ to arrive at
\be\begin{split}
&\partial_{\eta} \sigma^{ij}+2\HH\sigma^{ij}+v^k \partial_k \sigma^{ij}\\&+\sigma^{ik}\partial_k v^j+\sigma^{jk}\partial_k v^i=\frac{1}{\rho}\partial _k\left(\rho \sigma^{ijk}\right)\,,
\end{split}\ee
where on the right hand side we have introduced the third moment of the distribution function,
\be
 \sigma^{ijk} \equiv \langle u^iu^ju^k\rangle_p \,.
\ee
To truncate the collisionless Boltzmann hierarchy we neglect this term, setting 
$\sigma^{ijk}=0$. This is motivated by the fact that this tensor  has a vanishing background value and it contains an additional term $p/m \ll 1$ for non relativistic particles. In this way, we close the hierarchy, obtaining the system
\beqa \label{e:E0}
&0=&\!\!\pd_\eta\de + \nabla_\bx((1+\de)\bv)  \,, \\
&0=&\!\!\left(\pd_\eta +v^i\pd_i\right)v_j+\HH v_j  +\pd_i\Phi +\frac{1}{\rho}\pd_i(\rho\si_{ij}) \,, \label{e:E1}\\
&0=& \!\!\!(\partial_{\eta} +v^k \partial_k)\sigma^{ij}+2\HH\sigma^{ij}+\sigma^{ik}\partial_k v^j +\sigma^{jk}\partial_k v^i  \,, \qquad  \label{e:E2}
\eeqa
where $\Phi$ is determined self consistently by the Poisson equation~\eqref{e:Poisson}.

Taking the curl of eq.~\eqref{e:E1} we obtain the evolution equation for the vorticity, $\bom(\eta, \bx) \equiv \nabla_\bx \wedge \bv$. In vector notation
\be\label{e:vorticity}
\frac{\partial \bom}{d\eta}+\HH \bom-\nabla_\bx \wedge \left[\bv \wedge \bom\right]=-\nabla_\bx \wedge \left(\frac{1}{\rho} \nabla_\bx \left(\rho \sigma\right)\right) \,.
\ee
Here we have introduced the notation $(\nabla_\bx (\rho\sigma))^i\equiv \partial_j (\rho\sigma^{ji})$.  Clearly, if the velocity dispersion vanishes, $\si\equiv 0$, eq.~\eqref{e:vorticity} has no source term. This means that,  if the vorticity vanishes initially, it will remain zero for all times in accordance with Helmholz's theorem. 

\subsection{The Lagrangian equations}
The equations derived above are the fluid equations in `Eulerian space', i.e. $\bx$ denotes a fixed position in space. One can also write the equations in so called `Lagrangian' space where one follows a fluid element which is labelled by its initial position, see~\cite{Ehlers:1996wg} for a review of the Lagrangian approach.
The Lagrangian approach with non-vanishing velocity dispersion is also discussed in \cite{Aviles:2015osc,Morita:2001qe,Tatekawa:2002gf,Adler:1998dc,Buchert:1997dr}.

In Lagrangian coordinates, the position of a fluid element which was initially at position $\bq$ and which has velocity $\bu$ is
\be
   \bx = \bq +\bss(\eta, \bq,\bu)\,, 
\ee
where $\bss$ is called the Lagrangian map. The peculiar velocity of a fluid element is given by the implicit equation
\be\label{e:uimpli}
\bu(\eta,\bx) \equiv \frac{d\bx}{d\eta} =  \frac{d\bss}{d\eta}(\eta, \bq,\bu)\,. 
\ee
If there is no velocity dispersion, i.e. $f$ is a delta function in velocity space, for fixed $\eta$ and $\bx$, eq.~\eqref{e:uimpli} has a unique solution and the $\bu$-dependence of $\bss$ can be dropped. Velocity dispersion induces stochasticity in the velocity of a particle at position $\bx$. This is similar to the presence of multiple streams after shell crossing, with the difference that shell crossing occurs only when density perturbations are large and it is not stochastic, while velocity dispersion does not require large density perturbations and has a stochastic interpretation. The effect of velocity dispersion is certainly more relevant for warm and hot dark matter and related to the treatment of Refs.~\cite{Dupuy:2014vea,Dupuy:2015ega}, but we expect it to be present also to some extent for cold dark matter.

The standard Lagrangian displacement field is obtained by averaging $\bss$ over momenta,
\be\label{e:disp}
\frac{\partial\bpsi}{\partial \eta}(\eta, \bq)\equiv \left\langle\frac{\partial\bss}{\partial \eta}(\eta, \bq)\right\rangle_p=\bv\,.
\ee
The difference between $\bss$ and $\bpsi$ is related to the velocity dispersion via
 \be\nonumber
\bGamma(\eta, \bq,\bu)\equiv \bss-\bpsi\,, \quad \mbox{and }\quad 
\si^{ij}(\bq,\eta) = \langle\dot\Ga^i\dot\Ga^j\rangle_p \,,
\label{e:ga-si}
\ee
where $\bGamma$ is the stochastic piece of the displacement field and we have introduced the convective derivative along integral curves of the \emph{mean} velocity field. On a quantity $Y$:
\be
  \dot Y = \pd_\eta Y|_{\bq}= \pd_\eta Y|_{\bx} +\bv\nabla_{\bx}Y \,.
\ee
In Lagrangian coordinates this is just the partial time derivatives while in Eulerian coordinates the motion of the flow has to be accounted for.
We assume that we are still in the perturbative regime so that for a given velocity $\bu$ the application $\bq \ra \bx = \bq +\bss$ is invertible. We introduce its Jacobian,
\be\label{e:jacobian}\nonumber
J_{ij}\equiv \frac{\partial x^i}{\partial q^j}=\delta_{ij}+\frac{\partial \mathcal{S}^i}{\partial q^j} = \delta_{ij}+W_{ij}\,, \quad J\equiv \det (J_{ij})\,.
\ee
For later convenience we introduce also
\beqa
J  &\equiv& \text{det} (J_{ij}) = \frac{1}{6} \ep_{ijk}\ep_{pqr} J_{ip}J_{jq} J_{kr}\,,\\
(J^{-1})_{ij} &\equiv& \frac{1}{2J} \ep_{jkp}\ep_{iqr}J_{kq}J_{pr}\,, \label{jm1}\\
W^c_{ki} &\equiv&  \frac{1}{2} \ep_{kjp}\ep_{iqr}\mathcal{S}_{j,q} \mathcal{S}_{p,r}\,,
\eeqa
where all these functions are understood as functions of $\eta$ and $\bq$ for fixed $\bu$ and commas indicate partial derivatives with respect to $\bq$.

In order to transform the Eulerian derivatives of the previous section into Lagrangian ones we use
\be
\frac{\partial }{\partial \bx}=\frac{\partial \bq}{\partial \bx}\frac{\partial }{\partial \bq}=\bJ^{-1}\frac{\partial }{\partial \bq} \,.
\label{Eltrans}
\ee
We can write $J_{ij}$ as a deterministic part given by $\Psi$ and a stochastic part. The deterministic part is
\be\label{jacobianpsi}
\bar J_{ij}\equiv \delta_{ij}+\frac{\partial \Psi^i}{\partial q^j}\,, \qquad \bar J\equiv \det (\bar J_{ij})\,,
\ee
so that $J_{ij} = \bar{J}_{ij} + \Gamma_{i,j}$. In~\cite{Aviles:2015osc} it has been shown that the stochastic part of the transformation is very small and we therefore neglect it in the following. We will check a posteriori the consistency of this assumption. 
              
Using the above transformation with the deterministic part of $J$, which implies e.g. $\rho(\eta,\bx) =\rho(\bq)/J(\eta,\bq)$ we can transform our system~(\ref{e:E0}-\ref{e:E2}) to  Lagrangian coordinates. A lengthy calculation yields (see Appendix~\ref{a:fluid})\small
\begin{align}
&\left (\opt - 4 \pi G a^2 \bar \rho \right) \nabla \cdot \bpsi + \ep_{ijk} \ep_{ipq} \Psi_{j,p} \left ( \opt - 2 \pi G  a^2 \bar \rho \right)\Psi_{k,q} +\quad \nn\\
& + \ep_{ijk}\ep_{pqr} \Psi_{i,p}\Psi_{j,q}\left (\opt - \frac{4 \pi G a^2}{3} \bar \rho \right) \Psi_{k,r} = S_{\text{div}} \,,
\label{EulerDiv}\\
&\opt \left(\nabla \we \bpsi \right)_j - \left(\nabla \Psi_k \we \opt \, \nabla \Psi_k  \right)_j = \left(S_{\text{curl}}\right)_j \,,
\label{EulerCurl}\\
&\dot\sigma_{ij}+2 \mathcal{H}\sigma_{ij} = (S_\sigma)_{ij} \,,
\label{SigmaEq}
\end{align} \normalsize
where, following \cite{Matsubara:2015ipa}, we have introduced the differential operator $\opt= \partial^2_\eta + \HH \partial_\eta$; all time derivatives are at $\bq=$ constant, i.e. in Lagrangian space.
We have also split the Euler equation into its divergence part, \eqref{EulerDiv} and its curl part, \eqref{EulerCurl}. The sources on the right hand side are given by\small
\begin{align}
&S_{\text{div}} \!=\!-\frac{1}{4} \ep_{jsp}\ep_{iqr}\ep_{kab}\ep_{cdl} \bar J_{sq}\bar J_{pr} \nabla_i \!\!\left(\bar J_{ad}\bar  J_{bl} \nabla_c \!\left( \frac{\sigma_{kj}}{\bar J} \right) \right) \!+\nn\\
& \qquad+\text{[s.t.]}
\label{divsource} \,, \qquad\\
&\left(S_{\text{curl}}\right)_j\!=\!-\frac{1}{2} \ep_{jpq}\ep_{slr}\ep_{mnc}\bar  J_{kq} \nabla_p \left(\bar J_{nl} \bar J_{cr} \nabla_s \left( \frac{\sigma_{mk}}{\bar J} \right) \right) +\nn\\
&\qquad+ \text{[s.t.]} \,,
\label{curlsource}\\
&(S_\sigma)_{ij} = -\frac{1}{2 \,\bar J} \ep_{kbl}\ep_{pcq} \bar J_{lq} \bar J_{bc} \sigma_{ki} \nabla_{p} \dot \Psi_j + (i \leftrightarrow j) +\nn\\
&\qquad+ \text{[s.t.]}\,.
\label{sigmasource}
\end{align}\normalsize
The addition $\text{[s.t.]}$ indicates the stochastic part arising from the transformation (\ref{Eltrans}), which we will neglect in the following. Eqs.~(\ref{EulerDiv}) and (\ref{EulerCurl}) are the Lagrangian equivalent of the Euler equation, while eq.~(\ref{SigmaEq}) corresponds to the evolution equation for the velocity dispersion tensor. It is interesting to note that the continuity equation is automatically implemented in the Lagrangian formalism, i.e. the density contrast $\delta$ is not a dynamical variable in this frame, see Appendix~\ref{a:fluid} for details. For this reason, Lagrangian perturbation theory can explore further in the non-linear regime: a small perturbation of the Lagrangian path can carry non-linear information  about the corresponding Eulerian perturbation variables. 

It is possible to write an expression for the (Eulerian) vorticity $\bom=(\nabla_\bx \wedge \bv)$, as a function of the Lagrangian displacement field
\be\label{e:eulvort}
\omega_\ell  = \frac{a(t)}{J} \ep_{\ell kj} \left[ (1+ \mathcal{S}_{p,p}) \de_{ik} -W_{ik} +W_{ki}^c \right] \dot \Psi_{j,i}\,.
\ee
This result has also been obtained in Ref.~\cite{RRampf} in the form of an irrotationality condition for the fluid. If we neglect the stochastic terms it reduces to
$$
\omega_\ell = \frac{a(t)}{\bar J} \ep_{ijk} \bar J_{\ell i} \bar J_{mk} \dot{\bar{J}}_{mj}\,,
$$
in agreement with~\cite{Rampf:2016wom}. In Lagrangian coordinates the vorticity evolution equation~\eqref{e:vorticity}  becomes
\be\label{e:eulvort-evo}
\pd_{\eta} \omega_\ell + \HH \omega_\ell =\left(S_{\omega}^A\right)_{\ell}+\left(S_{\omega}^B\right)_{\ell}\,,
\ee
where the terms on the right hand side are given by\small
\begin{align}
\left(S_{\omega}^A\right)_{\ell} &\equiv  -\frac{1}{\bar J} \epsilon_{ijk}\epsilon_{pqr}\bar J_{iq}\bar J_{\ell r}\nabla_{p}\left(\dot{\Psi}_j\omega_k\right)+ \text{[s.t.]}\,, \label{e:sourceA-bom} \\
\left(S_{\omega}^B\right)_{\ell} &\equiv  \frac{1}{2 \bar J}\epsilon_{jab}\epsilon_{pqr}\epsilon_{nms}\bar  J_{\ell q} \bar J_{ir} \nabla_p\left[ \bar J_{am} \bar J_{bs} \nabla_n\left(\frac{\sigma^{ij}}{\bar J}\right)\right]+\nn\\
&+\text{[s.t.]}\,. \label{e:sourceB-bom}
\end{align}\normalsize
Again [s.t.] indicates a stochastic component which we do not explicitly write. Here $S_{\omega}^A$ is a homogeneous term, however  in the perturbative treatment outlined below, $S_{\omega}^A$ of order $n$ will only contain $\bom$ and $\dot\Psi$ of order $n-1$ and lower. Therefore it can be regarded as a source term of the $n$th order perturbations. The term $S_{\omega}^B$ is a source term proportional to the velocity dispersion.
As we have already seen in the Eulerian approach, if $\si_{ij}=0$, $\bom\equiv 0$ solves the vorticity equation, i.e.  velocity dispersion is needed to source vorticity if we assume the velocity distribution of dark matter after inflation to be curl free. In other words the $n$th order source term $(S_{\omega}^A)^{(n)}$ can have a non zero value only if $(S_{\omega}^B)^{(n-1)} \neq 0$.

\subsection{Lagrangian perturbation equations}

The equations of the previous section are exact and valid as long as $J\equiv\det \bJ\neq 0$, i.e. as long as the map from Lagrangian to Eulerian coordinates is invertible, which means that there has not been any shell crossing. Here we analyse the system (\ref{EulerDiv} --\ref{e:sourceB-bom}) making use of Lagrangian perturbation theory (LPT).
We introduce a perturbative expansion for the displacement field, the velocity dispersion and the vorticity,
\be\label{e:pert} \nonumber
\bpsi = \sum\limits_{n=1}^{\infty} \bpsi^{(n)}\,,\qquad \sigma_{ij} = \sum\limits_{n=0}^{\infty} \sigma^{(n)}_{ij}\, \qquad \bom = \sum\limits_{n=1}^{\infty} \bom^{(n)}\,.
\ee
We emphasize that the vorticity is not an independent  variable. Indeed, 
Eq.~\eqref{e:eulvort} implies
\begin{widetext}
\begin{align}
\omega_\ell^{(n)} &= \,\, a(t) \sum\limits_{\al+\b=n} \left( \frac{1}{ J}\right)^{(\al)}  \ep_{\ell kj}\dot \Psi_{j,k}^{(\b)}+ a(t)  \sum\limits_{\al+\b+\gamma=n} \left( \frac{1}{J}\right)^{(\al)} \ep_{\ell kj} \left( \Psi_{p,p}^{(\b)} \dot \Psi_{j,k}^{(\ga)}-  \Psi_{i,k}^{(\b)} \dot \Psi_{j,i}^{(\ga)} \right)+\nn\\
&\qquad\qquad  + a(t)  \sum\limits_{\al+\b+\ga+\de=n} \left( \frac{1}{J}\right)^{(\al)}  \ep_{iqr}   \Psi_{j,q}^{(\b)} \dot \Psi_{j,i}^{(\ga)} \Psi_{\ell,r}^{(\de)}\,.\label{vortexp}
\end{align}
The inverse of the determinant of the Jacobian up to third order can be written as
 (see Appendix~\ref{J} for the expansion up to a generic order $n$) \be \begin{split}
\frac{1}{J}& \simeq  \,\,1- \trr[W^{(1)}] +\frac{1}{2} \trr[W^{(1)}]^2 +\frac{1}{2} \trr[(W^{(1)})^2] - \trr[W^{(2)}] -\frac{1}{6} \trr[W^{(1)}]^3 -\frac{1}{3} \trr[(W^{(1)})^3] \\ &-\frac{1}{2} \trr[W^{(1)}] \trr [(W^{(1)})^2] +\trr[W^{(1)}W^{(2)}] + \trr[W^{(1)}] \trr[W^{(2)}] - \trr[W^{(3)}] + \cdots
\label{det3ord}\end{split}
\ee
As in standard Lagrangian perturbation theory, we write the evolution equations as differential equations in the 'time' variable $\tau =\log D_+=\log a$. Here $D_+$ is the growing mode of linear density perturbations and the last equal sign is valid only in
an Einstein-de Sitter Universe (EdS). Other backgrounds request factors of $f$ and $f^2$ in the equations, where 
$f=d\log D_+/d\log a \simeq \Om_m^{0.56}$ for a $\La$CDM  Universe.
In terms of this variable, and during  matter domination, the perturbation expansion of the evolution equations~\eqref{EulerDiv} to \eqref{SigmaEq} become  
\begin{eqnarray}
\left[ \frac{\pd^2}{\pd \tau^2} + \frac{1}{2} \frac{\pd}{\pd \tau} -\frac{3}{2} \right] \nabla \cdot \bpsi^{(n)} &=& - \sum\limits_{\al+\b=n} \ep_{ijk}\ep_{ipq} \Psi_{j,p}^{(\al)} \left[ \frac{\pd^2}{\pd \tau^2} + \frac{1}{2} \frac{\pd}{\pd \tau} -\frac{3}{4} \right] \Psi_{k,q}^{(\b)} \nn \\
&& -\frac{1}{2}  \sum\limits_{\al+\b+\ga=n} \ep_{ijk}\ep_{pqr} \Psi_{i,p}^{(\al)} \Psi_{j,q}^{(\b)} \left[ \frac{\pd^2}{\pd \tau^2} + \frac{1}{2} \frac{\pd}{\pd \tau} -\frac{1}{2} \right] \Psi_{k,r}^{(\b)}+\HH^{-2}S_{\text{div}}^{(n)}\,,  \qquad
\label{gradeqVDT}\\
 \left[ \frac{\pd^2}{\pd \tau^2} + \frac{1}{2} \frac{\pd}{\pd \tau} \right] \nabla \we \bpsi^{(n)} &=& \sum\limits_{\al+\b=n} \nabla \Psi_i^{(\al)} \we  \left[ \frac{\pd^2}{\pd \tau^2} + \frac{1}{2} \frac{\pd}{\pd \tau} \right] \nabla \Psi_i^{(\b)}+\HH^{-2}S_{\text{curl}}^{(n)}\,,
 \label{curleqVDT}\\
\HH \left[ \frac{\pd}{\pd \tau} +2 \right]\sigma^{(n)}_{ij}&=&S_{\sigma}^{(n)} 
\label{sigmaeqVDT} \,.
\end{eqnarray}
The source terms up to order $n=2$ for the displacement field and for the velocity dispersion are 
\beqa
S_{\text{div}}^{(1)} &=& \sigma^{(0)} \nabla^2\left( \Psi_{i,i}^{(1)} \right) - \nabla_i \nabla_j \sigma_{ij}^{(1)} \,,
\label{sourcediv1} \\
S_{\text{div}}^{(2)} &=& -  \Psi_{i,i}^{(1)}  \nabla_j \nabla_k \sigma^{(1)}_{jk} + \na_i \Psi_{j,j}^{(1)} \na_k \sigma^{(1)}_{ik} +2 \na_i \Psi^{(1)}_j  \na_k \na_j \sigma^{(1)}_{ik}  + \na_i \sigma^{(1)}_{jk} \na_k \na_j \Psi^{(1)}_i  + \sigma^{(1)}_{ij} \na_i \na_j \Psi_{k,k}^{(1)} \nn\\
&&+\sigma^{(0)} \left[\na^2 \Psi_{i,i}^{(2)} -2 \na_i \na_j \Psi_{k,k}^{(1)} \na_i \Psi_j^{(1)}- \na^2 \Psi^{(1)}_i \na_i \Psi^{(1)}_{j,j} -\na_i \Psi^{(1)}_j \na^2 \na_j \Psi^{(1)}_i\right.\nn\\
&&\left. - \na_i \na_j \Psi^{(1)}_k \na_i \na_k \Psi_j^{(1)}+ \Psi_{i,i}^{(1)} \nabla^2 \Psi_{r,r}^{(1)} \right]- \nabla_i \nabla_j \sigma^{(2)}_{ij} \,,
\label{sourcediv2}\\
\left(S_{\text{curl}}^{(1)}\right)_j &=& \ep_{jik} \na_k \na_r \sigma^{(1)}_{ri}  \,,
\label{sourcecurl1}\\
\left(S_{\text{curl}}^{(2)}\right)_j &=& \ep_{jkp} \Bigl[ \na_p \na_i \sigma^{(2)}_{ik} - \sigma^{(1)}_{k\ell}  \na_p \na_\ell \Psi_{r,r}^{(1)} + \na_i \sigma^{(1)}_{\ell p} \na_k \na_\ell \Psi^{(1)}_i - \na_\ell\Psi_{i,i}^{(1)} \na_p \sigma_{\ell k}^{(1)} \nn\\
&&- \na_\ell \Psi^{(1)}_i \na_p \na_i \sigma^{(1)}_{\ell k} + \na_k \Psi_i^{(1)} \na_p \na_m \sigma^{(1)}_{im} \Bigr] \,,
\label{sourcecurl2}\\
\left(S_{\sigma}^{(1)}\right)_{ij} &=& - \HH \sigma^{(0)} \left( \na_i \Psi'^{(1)}_j + \na_j \Psi'^{(1)}_i \right)  \,,
\label{sourceVDT1}\\
 \left(S_{\sigma}^{(2)}\right)_{ij} &=& - \HH\sigma^{(1)}_{ik} \na_k \Psi'^{(1)}_j - \HH \sigma^{(0)} \left(\Psi'^{(2)}_{i,j} - \na_k \Psi_i'^{(1)} \na_j \Psi_k^{(1)} \right) + (i \leftrightarrow j)  \,.
\label{sourceVDT2}
\eeqa \end{widetext}
The primes appearing in the source terms indicate derivatives wrt $\tau$.

The only quantity which does not vanish at order $n=0$ is the velocity dispersion for which \eqref{sigmaeqVDT} implies
\be
\si^{(0)}_{ij} = \frac{\si_0}{3}a^{-2}\de_{ij} = \si^{(0)}\de_{ij} \,.
\ee
With the normalisation of the present scale factor, $\si_0/3$ is the present dark matter pressure. This parameter $\si_0$ is determined by the dark matter properties and it is clearly much larger for warm dark matter than for standard cold dark matter.

It would be possible to solve these equations for the displacement  field and the VDT tensor and then calculate the vorticity using \eqref{vortexp}. However, since we are mainly interested in the vorticity, it is simpler to directly solve the vorticity equation (\ref{e:eulvort-evo}) which in terms of $\tau=\log a$ becomes
\be\label{e:master}
\HH \left[\frac{\partial}{\partial \tau} \bom^{(n)}+\bom^{(n)} \right]=S_{\omega}^{A\,(n)}+S_{\omega}^{B\,(n)}\,.
\ee
with source terms\begin{widetext}
\beqa
S_{\omega}^{A\,(1)} &=& 0  \,,
\label{sourceomegaA01}
\\
 \left( S_{\omega}^{A\,(2)} \right)_\ell &=& \HH\left[\Psi'^{(1)}_\ell \na_j \omega^{(1)}_j -\omega^{(1)}_\ell \na_j \Psi'^{(1)}_j +\omega^{(1)}_j \na_j \Psi'^{(1)}_\ell -\Psi'^{(1)}_j \na_j \omega^{(1)}_\ell \right]\,,
\label{sourceomegaA2}
\\
\left( S_{\omega}^{B\,(1)} \right)_\ell &=& \ep_{\ell pr} \na_r \na_j \sigma^{(1)}_{jp} \,,
\label{sourceomegaB1}\\
\left( S_{\omega}^{B\,(2)} \right)_\ell &=& \ep_{\ell ij} \left(\na_j\na_b \sigma^{(2)}_{bi} -  \Psi^{(1)}_{p,p} \na_j \na_b \sigma^{(1)}_{bi}\right) - \ep_{\ell bj} \sigma^{(1)}_{bi} \na_j\na_i \Psi^{(1)}_{p,p} + \ep_{bjp} \na_b \Psi^{(1)}_\ell \na_p \na_i \sigma^{(1)}_{ij}  \nn\\
&&+ \ep_{\ell jp} \left(\na_b \sigma^{(1)}_{ip} \na_j \na_i \Psi^{(1)}_b - \na_i  \Psi^{(1)}_{b,b} \na_p \sigma^{(1)}_{ij} - \na_i \Psi^{(1)}_b \na_p \na_b \sigma^{(1)}_{ij} \right) + \ep_{\ell ip} \na_i \Psi^{(1)}_b \na_p \na_j \sigma^{(1)}_{bj}  \,.
\label{sourceomegaB2}
\eeqa\end{widetext}
Here $S_{\omega}^{A\,(1)} = 0$ because background vorticity $\omega^{(0)}$ is not allowed by isotropy. 

The goal of this work is to study the vorticity which is induced by velocity dispersion. In the presence of velocity dispersion we can write the displacement field as
\be
\Psi=\Psi_{\text{st}}+\delta \Psi_{\sigma}\,,
\ee
where  $\Psi_{\text{st}}$ denotes the standard LPT result and  $\delta \Psi_{\sigma}$ the correction induced by the coupling to the VDT given by the source terms of Eqs.~\eqref{gradeqVDT} and \eqref{curleqVDT}. The solutions to the perturbation equations for $\Psi^{(n)}$ up to third order for vanishing velocity dispersion are well known and we do not repeat them here. Since we need them later, we present the expressions in Appendix~\ref{a:pert}. Plugging in the sources on the right hand side of eqs.~(\ref{EulerDiv}) and (\ref{EulerCurl})  the standard LPT result  for the displacement, we see that the correction $\delta \Psi_{\sigma}$ induced by the coupling to the VDT is always subleading with respect to $\Psi_{\text{st}}$.  Moreover the VDT solution introduces the small background constant $\sigma_0$ which characterize the DM pressure and further suppresses this correction. At first order, for example, we  obtain
\be
\Psi_{\text{st}}^{(1)} \propto D_+ \,\,\,\,\, , \,\,\,\,\, \delta \Psi_{\sigma}^{(1)} \propto \sigma_0 D_+^{-2} \,.
\ee
Therefore, as long as we want to keep only the leading growing mode of $\Psi$, we can just use in the source of equation  (\ref{SigmaEq}) the standard LPT result and neglect $\delta \Psi_{\sigma}$.

Solving \eqref{sigmaeqVDT} in this approximation in Fourier space, we obtain for the fastest growing (least decaying) modes, 
\beqa
\sigma_{ij}^{(1)}(\bk,t) &=& \frac{2}{3} \frac{k_i k_j}{k^2} \sigma_0 D_+^{-1} \de_0(\bk)\,,
\label{Fsigma1}\\
\sigma_{ij}^{(2)}(\bk,t) &=& \frac{1}{3} \sigma_0  \,  \mathcal{I}_{ij} (\bk)\,.
\label{Fsigma2}
\eeqa
Here $\de_0(\bk) $ is the linear density fluctuation normalized to one today (in Fourier space) and we have introduced
\be
\mathcal{I}_{ij} (\bk) =\frac{1}{2}\, \int \frac{\dd^3 w}{(2 \pi)^3} \, \, K^{(2)}_{ij} (\bw,\bk-\bw) \de_0(\bw) \de_0(\bk-\bw)\,.
\ee\normalsize
The kernel $K^{(2)}_{ij}$ is  given by 
\begin{align}
K^{(2)}_{ij}(\bk_1, \bk_2) = &(\bk_1 \cdot \bk_2) L^{(1)}_i(\bk_1)L^{(1)}_j(\bk_2) +\nn\\
&+ k_i L^{(2)}_j (\bk_1,\bk_2) + (i \leftrightarrow j) \,,
\label{kernelK2}
\end{align}
\normalsize with
\begin{align}
&L^{(1)}(\bk) = \frac{\bk}{k^2} \,,\nn\\
& L^{(2)}(\bk_1, \bk_2)= \frac{3}{7} \frac{\bk}{k^2} \left[1- \left(\frac{\bk_1 \cdot \bk_2}{k_1 k_2} \right)  \right] \,.
\end{align}
These results are in agreement with~\cite{Aviles:2015osc}.

Note that we have not added a homogeneous solution to the higher order terms $\sigma^{(n)}$, $n\geq 1$ as we assume $\si^{(0)}$ to represent the full homogeneous mode.

We can infer the  time evolution of the fastest growing mode of $\si_{ij}^{(n)}$ when neglecting the stochastic part. Since  the fastest growing mode of its source $S^{(n)}_\si$ behaves like  $S^{(n)}_\si\propto \HH \si^{(0)} \bpsi^{(n)} \propto D_+^{n-5/2}$, 
we obtain $\si_{ij}^{(n)} \propto \langle\dot\Ga_i\dot\Ga_j\rangle_u^{(n)} \propto \si_0D_+^{n-2}$
and $\Ga^{(n)}_i \propto \sqrt{\si_0}D_+^{n-1/2}$, while we know $\bpsi^{(n)} \propto D_+^{n}$. This means that every time we have neglected  a term containing $\Gamma_{k,j}$ in the sources we have considered an identical term in $\Psi_{k,j}$ which grows faster and it is not suppressed by the factor $\sqrt{\sigma_0}$. Hence for sufficiently small $\si_0$ it is justified to neglect the stochastic contribution.

\section{Results}\label{s:res}

Using the results for $\si^{(1)}$ and $\si^{(2)}$, we can infer the time evolution for the fastest growing contribution to the vorticity source term, $S_{\omega}^{A\,(n)} \propto \bom^{(n-1)} D_+$ and $S_{\omega}^{B\,(n)} \propto D_+^{n-2}$  such that
\be
\pd_\eta \bom^{(n)} + \HH \bom^{n)} \simeq \bom^{(n-1)} D_+  + D_+^{n-2} \,.
\ee
At first order we have  $S^{A\,(1)}_\omega=0$ and inserting \eqref{Fsigma1} we obtain in Fourier space
\be
\left( S_{\omega}^{B\,(1)} \right)_\ell \propto \ep_{\ell pr} k_p k_r  \,  \de_0(\bk) =0  \,,
\ee
hence $\bom^{(1)}=0$. This implies that also at second order $S^{A\,(2)}_\omega\propto \omega^{(1)} =0$. But the source term $S_{\omega}^{B\,(2)}$ is non-vanishing and, since  $S^{B\,(n)}_\omega\propto D_+^{n-2}$, it is constant at second order. Therefore, already at second order we obtain a growing mode: $\bom^{(2)} \propto \sqrt{D_+}$. At higher orders the source term $S^{A\,(n)}_\omega\propto  \bom^{(n-1)} D_+ $ leads to the faster growing mode
which yields $\bom^{(n)} \propto D_+^{(3n-5)/2}$ for $n\geq 2$. 

More precisely, the second order source term  $S_{\omega}^{B\,(2)}$ is \begin{widetext}
 \be
 \left(S_{\omega}^{B\,(2)} \right)_\ell = -\frac{\sigma_0}{3} \ep_{\ell ij} k_j k_b \, \mathcal{I}_{bi} (\bk) = \sigma_0 \int \frac{\dd^3 \bw}{(2 \pi)^3} \frac{\bw \cdot (\bk-\bw)}{w^2 |\bk-\bw|^2} \bk \cdot \bw \left(\bw \we \bk \right)_\ell  \de_0(\bw) \de_0 (\bk-\bw)  \,,
\ee
so that 
\be\label{e:vort2}
\bom^{(2)} (\bk,\eta) = \frac{2}{3} \frac{\sigma_0}{\HH_0 \sqrt{ \Omega_m}} \sqrt{D_+} \int \frac{\dd^3 \bw}{(2 \pi)^3}\, \frac{\bw \cdot (\bk-\bw)}{w^2 |\bk-\bw|^2} \,(\bk \cdot \bw)\, \left(\bw \we \bk \right)  \de_0(\bw) \de_0 (\bk-\bw) \,.
\ee\end{widetext}
Eq~\eqref{e:vort2} is our main result. In the following we compute and discuss its power spectrum.

The power spectrum of $\bom^{(2)}$ is defined by
\begin{align}\label{e:pow-def}
\langle &\omega_i^{(2)} (\bk,\eta) \omega_j^{(2)\,*} (\bk',\eta)\rangle =\nn\\
&=(2\pi)^3 \left(\de_{ij}-\hat k_i\hat k_j\right) \de(\bk-\bk')P_\omega(k,\eta) \,.
\end{align}
Here $\hat\bk =\bk/k$, where $k=|\bk|$ and the pre factor is a consequence of the fact that  vorticity is divergence free, $\bom\cdot\bk=0$. This can also be checked directly from the solution \eqref{e:vort2} of $\bom$.  Inserting the result \eqref{e:vort2} in the above definition we find\begin{widetext}
\be\label{e:pow-def2}
P_\omega(k) =\frac{1}{9} \frac{\sigma_0^2 D_+(\eta)}{\HH_0^2 \Omega_m} \int \frac{\dd^3 \bw}{(2 \pi)^3} \left(\frac{\bw \cdot (\bk-\bw)}{w^2 |\bk-\bw|^2}\right)^2
\left|\bw \we \bk \right|^2 \left[2 \bk \cdot \bw -k^2 \right]^2 P_\de(w) P_\de (|\bk-\bw|) \,,
\ee \end{widetext}
where we have introduced the linear density power spectrum, $\langle \de_0(\bk) \de_0^* (\bk') \rangle= (2\pi)^3 P_\de (k) \de(\bk-\bk')$.

\begin{figure*}
\noindent\makebox[\textwidth]{\includegraphics[width=1\textwidth]{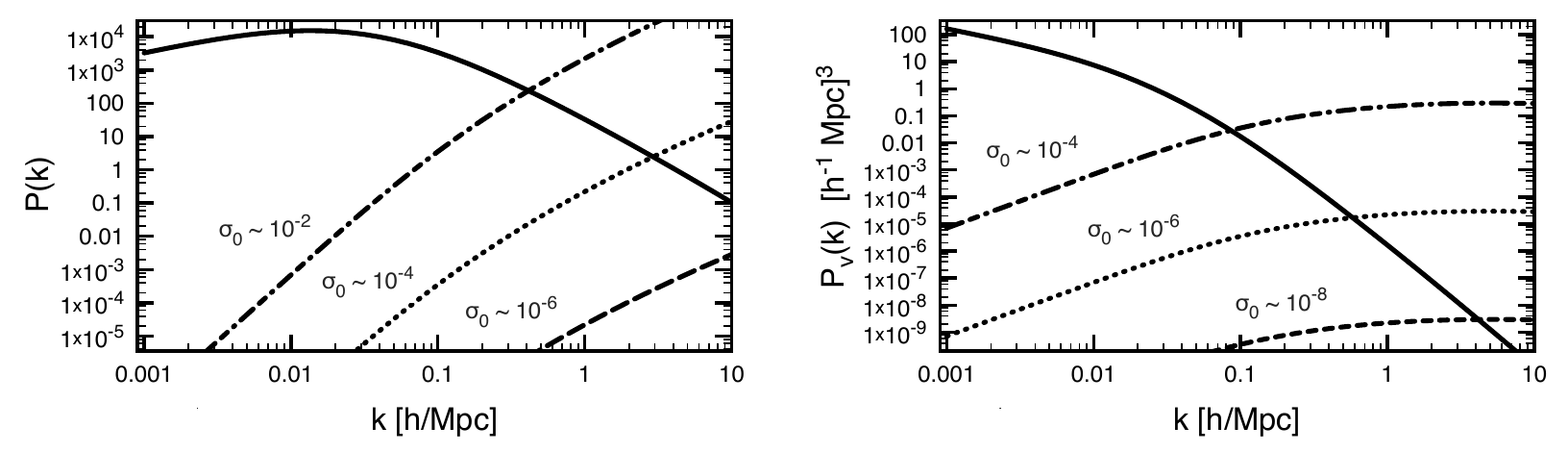}}
\caption{(\emph{Left}) We plot the matter power spectrum $P_\delta(k)$ at z=0 (solid) and the vorticity power spectrum $P_\omega(k)$ for different values of $\si_0$ [in units of 
$(h^{-1}$Mpc$)^3$ and $(h^{-1}$Mpc$)^5$ respectively] . (\emph{Right}) We also plot the power spectrum of $P_{v_R}(k)$ (dashed), i.e. the rotational component of the peculiar velocity, which is given by $k^{-2}P_\omega(k)$ and compare it with the gradient velocity power spectrum (solid), $P_v(k)=\Om_m^{1.14}(\HH/k)^2P_\delta(k)$. } 
\label{fig:powerspectra}
\end{figure*}
We plot $P_\omega$ together with $P_\de$ in Fig.~\ref{fig:powerspectra} for different values of $\si_0$. Since $\bom (k) = i \bk \we \bv(k)$ we can trivially compute the rotational component of the peculiar velocity given by $\bv^R(k)=i k^{-2} \bk \we \bom(k)$. Its spectrum can be written in the form
\begin{align}
\langle &v_i^R (k,\eta) v_j^{R*}(k',\eta) \rangle =\nn\\
&=(2\pi)^3 \left(\de_{ij}-\hat k_i\hat k_j\right) \de(\bk-\bk')P_{v_R}(k,\eta) \,.
\end{align}
A short calculation shows that $P_{v_R}(k) = k^{-2} P_\omega(k)$.
We assume a power law for the inflationary power spectrum with spectral index $n_s=0.96$ so that the linear matter spectrum is given by 
\begin{align}
 P_\de (k)&\equiv P_\de (k,\eta_0) =\nn\\
 &= 2 \pi^3 k^{-3} \de_H^2 \left(\frac{k}{H_0} \right)^{3+n_s} T^2(k) \,.
\label{matterspectrum}
\end{align}
Planck normalisation requests  $\de_H \simeq 4.6 \times 10^{-5}$ and we use the  approximation by Eisenstein and Hu~\cite{Eisenstein:1997ik} for the transfer function $T(k)$.  The amplitude of the vorticity power spectrum depends quadratically   on the velocity dispersion $\si_0$. For dark matter particles which are non-relativistic at the moment of decoupling, $t_*$, following a Maxwell-Boltzmann distribution, the velocity dispersion is given by 
\be 
\si^{(0)} = 3\frac{T_*}{m}\left(\frac{a_*}{a} \right)^2 \,, \ee
which implies $\si_0 = 3 \, (T_* /m)(1+z_*)^{-2}$. Hence $\si_0$ is proportional to the present dark matter `temperature'\footnote{Strictly speaking the the word `temperature' is somewhat abusive after decoupling as dark matter is no longer in thermal equilibrium. Nevertheless, its momenta are just redshifted so that the momentum distribution remains a Maxwell Boltzmann distribution if we rescale the temperature with $a^{-2}$. Strictly speaking, however this $T$ is no longer a thermodynamical temperature but simply a time dependent parameter of the distribution function.} $T_0=T_*(1+z_*)^{-2}$  and it is higher if we consider a DM model with an higher $T_0$.  For extremely  relativistic particles with $\si_0\sim \langle u^2\rangle_p \sim 1$, hot dark matter (HDM) our approximation schemes breaks down since we have neglected additional powers of $\sqrt{\si_0}$ which is not justified in this case. Furthermore, for relativistic particles, higher terms in the Boltzmann hierarchy cannot be neglected, as we have done here by setting $\si_{ijk} =0$. For warm dark matter (WDM)  typical  decoupling velocities are still relativistic and the distribution function at decoupling is a function of $\exp(-P_{\text{phys}}/T)$ so that redshift of momenta leads to a significantly larger 'effective temperature', $T_0=T_*/(1+z_*)$ and  also larger velocity dispersion, see Refs.~\cite{Kunz:2016yqy,Thomas:2016iav} for recent discussions of dark matter properties.

\section{Conclusions}\label{s:con}

In the usual treatment, cold dark matter is described as a perfect fluid, fully characterized by its density fluctuation $\de$ and the divergence of the  velocity field, $\theta$. Mathematically, this implies that the evolution of dark matter is governed by the first two moments of the Boltzmann hierarchy: the velocity dispersion and higher moments of the distribution function are set exactly to zero.  However we know that this is an often sufficient but not realistic description, since dark matter species have a finite temperature  and therefore a non-vanishing velocity dispersion. The perfect fluid approximation is  expected to be a rather good description as long as cold dark matter species are considered, in the single streaming regime. On the other hand, this approach is not expected to capture important features of warm dark matter clustering, since in this case velocity dispersion might play an important role.

In this paper we propose to generalize the standard description of dark matter as a perfect fluid by truncating the Boltzmann hierarchy at the second moment and by including velocity dispersion in the field equations.  We explicitly compute the velocity dispersion tensor up to second order in perturbation theory, we deduce its  time evolution at every order and we find that growing modes start from the third order. Our results depend only on the linear matter density field and therefore can be applied to any other perturbation scheme. Our findings for the velocity dispersion tensor, agree with the ones of  \cite{Aviles:2015osc}, where the effects of velocity dispersion on the velocity divergence power spectrum have been studied. We have computed  the vorticity induced by the presence of velocity dispersion up to second order in Lagrangian perturbation theory in a matter dominated Universe and we found that at second order, vorticity grows like the square root of the scale factor: as soon as primordial velocity dispersion is not exactly vanishing as we realistically expect, vorticity is generated in our universe. We have computed the corresponding power spectrum; given the small-scale behavior for the matter power spectrum $P_\de \propto k^{n_s-4} \text{Log}^2(k)$ the spectrum of the rotational part of the DM velocity has a slow small-scales decay $P_{v_R} \propto k^{n_s-2} \text{Log}^4(k)$. 

To our knowledge, vorticity of dark matter velocity has never been calculated in a perturbative approach and it will be very interesting to compare this result to N-body simulations with initial conditions including a non-vanishing  velocity dispersion. Recently it has been argued that if velocity dispersion is considered appropriately in the initial conditions of N-body simulations of warm dark matter, structure formation qualitatively changes and an interplay between a standard hierarchical top-down model at large scales and a bottom-up at small scales is observed~\cite{Paduroiu:2015jfa}. On the one hand we stress that our predicted generation of vorticity can not be observed in simulations with vanishing (or underestimated) velocity dispersion.  On the other hand every N-body code is afflicted by numerical dispersion and finite resolution which will induce an effective velocity dispersion as soon as there are more than one particle per grid cell. This induces velocity dispersion and therefore vorticity.  Furthermore, we do expect shell crossing to lead to the generation of vorticity. It is an unsolved problem to which amount  the vorticity field found in numerical simulations in~\cite{Pueblas:2008uv} is affected by finite resolution and which portion of it is physical and independent of resolution. (Note that even though the results of Ref.~\cite{Pueblas:2008uv} seem to converge for the two highest mass resolutions, both these simulations have the same softening lengths and therefore the same spatial resolution.)

Our results depend on the small dimensionless quantity, $\sigma_0$, the trace of the background velocity dispersion tensor evaluated at present time. This introduces a  free-streaming scale, below which dark matter clustering is suppressed, $\lambda_\text{fs}\sim \sigma_0/\mathcal{H_0}$ (see the extensive  treatment in~\cite{Aviles:2015osc}).  
If the dark matter is a thermal relic, $\sigma_0$ can be related to its temperature to mass ratio. However, since $\sigma_0$ is a small quantity, it is susceptible to non-negligible corrections from higher orders in perturbation theory, as well as contributions from higher moments in the Boltzmann hierarchy. 

The linear theory of CDM velocity dispersion, has been recently studied in \cite{Piattella:2015nda} suggesting an upper bound of about $\sigma_0\sim 10^{-14}$ for thermally produced dark matter. The warm dark matter scenario is different: here the DM particles are still relativistic at freeze-out, and velocity dispersion is important at the stage of structures formation, inducing a quite large free-streaming scales. We underline that the perturbation equations derived are valid only for  non-relativistic dark matter and our solutions can not be applied to HDM as long as the particles are still relativistic. Then, the traditional Vlasov perturbation scheme in which the $\delta f$ is considered as a small perturbation quantity applies. 
However, as soon as the HDM particles become non-relativistic, say $\si_0\lesssim 0.01$, the scheme presented here is applicable and simpler than a full Vlasov approach.
This is especially true for massive neutrinos: their linear velocity dispersion is relatively large, $\sigma_0\sim 10^{-7} eV^2/m_{\nu}^2$ (an accurate measurement of it represents a promising method to infer their absolute mass). Our treatment can be applied  also to neutrinos in the non-relativistic stage of their evolution. For more details see~\cite{Shoji:2010hm} and references therein. 

During the first stages of gravitational collapse, dark matter is single streaming. Later, non linear collapse makes different streams converge, leading to the formation of matter caustics and ultimately to shell crossing.  Our formalism can capture the mildly non linear regime, i.e. the regime in which the density contrast grows to a value of order unity. This is an advantage of Lagrangian perturbation theory over the standard (Eulerian)  one. However, at shell crossing both Eulerian perturbation theory and LPT break down and one has to rely on  N-body simulations.


It would be interesting to measure cosmological vorticity and its correlation function in more detail.
The work cited in the introduction~\cite{Taylor:2016rsd} used the correlation of angular momenta of radio galaxies. Another possibility might be redshift space distortions in the number counts coming from vortical motion~\cite{Durrer:2016jzq}.  It is difficult to see how the observed large scale vorticity correlation can come from non-linearities on small scales. On one hand, the relation between the dark matter clustering calculated here and the galaxies may be quite complicated (bias). On the other hand, it is possible that small scale vorticity, once it becomes non-linear, sweeps into larger scales by inverse cascade.  
  
\acknowledgments We thank Julian Adamek for helpful discussions. The tensorial algebra for the calculation of some sources has been partially performed with \texttt{xAct}~\cite{Martin-Garcia:2007bqa}. This work is supported by the Swiss National Science Foundation.
\newpage

\appendix

\section{Lagrangian fluid equations}\label{a:fluid}

In this appendix we present some details of the derivation of the Lagrangian equations (\ref{EulerDiv})-(\ref{SigmaEq}) starting from the system of equations (\ref{e:E0})-(\ref{e:E2}) (continuity, Euler and VDT equations), which describes the evolution of the fluid in Eulerian coordinates.  Most of this can also be found in the literature, e.g.~\cite{Buchert:1992ya,Ehlers:1996wg,Rampf:2012xx}.

The continuity equation can be rewritten as 
\be
d^3 \bx\,\rho(\eta, \bx)=d^3\bq\, \rho(\bq)\,,
\ee
or equivalently 
\be\label{cont}
\rho(\eta, \bx)=\rho(\bq)/ J(\eta, \bq)\,.
\ee
Neglecting stochastic contributions, it follows that\footnote{We are using the standard result $d\left(\det A\right)=\det A\, \text{tr} \left(A^{-1} dA\right)$. }
\be
\frac{d J}{d\eta}= J \,\text{Tr}\left(\bJ^{-1} \frac{d \bJ}{d \eta}\right)=J\, \nabla\cdot \bv\,.
\ee
Using (\ref{cont}) and the above result, we obtain
\be
0= \frac{\partial}{\partial\eta}\left(\rho\,J\right)=J\left(\frac{\partial\rho}{\partial \eta}+\rho \nabla \cdot \bv\right)\,,
\ee
and we recover the known fact that in the Lagrangian picture the continuity equation is automatically implemented, independently on the specific form of the map between Lagrangian and Eulerian coordinates. 

In the Lagrangian picture the density field is not anymore a dynamical variable. This is one of the great advantages of using Lagrangian perturbation theory over standard perturbation theory: beside having a dimensional reduction of the system \cite{Buchert:1992ya}, when we study perturbation theory in LPT, we do not need to linearize over the density field. It follows that 
LPT can explore further into the non-linear regime \cite{Buchert:1992ya}.

We now turn to eqs.~(\ref{e:E1})-(\ref{e:E2}). With the help of the convective derivative,
$\dot X\equiv \partial_\eta X +v^i\partial_{x^i}X$, they can be rewritten as two equations for the evolution of the displacement field and for the stress-tensor 
\be
\ddot{\Psi^i}(\bq)+\HH \dot{\Psi^i}(\bq)=-\frac{\partial}{\partial x^i}\Phi(\bq+\bpsi)-\frac{1}{\rho}\frac{\partial\left( \rho\, \sigma_{ij}\right)}{\partial x^j}\,,\label{EulerL}
\ee
\begin{align}
\dot{\sigma}^{ij}(\bq+\Psi)+2\HH &\sigma^{ij}+\dot{\Psi}^k \partial_k \sigma^{ij}+\nn\\
&+\sigma^{ik}\partial_k \dot{\Psi}^j+\sigma^{jk}\partial_k \dot{\Psi}^i =0\,.\label{newL}
\end{align}
We can write the divergence and curl part of Euler equation separately. For the divergence part we obtain eq.~(\ref{EulerDiv}) while  the curl part yields eq.~(\ref{EulerCurl}) in the main text. In eqs.~(\ref{EulerDiv}) and (\ref{EulerCurl}), the contribution from velocity dispersion to Euler equation is encoded in the sources $S_{\text{div}}$ and $S_{\text{curl}}$. Furthermore we have denoted the stochastic terms involving the field $\bGamma$ with [s.t.]. This is different from what it is usually done since, normally, the curl and divergence equations are the starting point to develop recursion relations. If we neglect velocity dispersion, the first term in the equations is linear in the displacement field while the other terms are higher order and they can be considered as sources (see \cite{Matsubara:2015ipa},\cite{Ehlers:1996wg},\cite{Rampf:2012xx} for details on recursion relations in standard LPT). We stress that if we include velocity dispersion, terms   linear in $\Psi$ may arise from the sources. Applying a similar procedure, the equation for the velocity dispersion (\ref{e:E2}) is written as eq.~(\ref{SigmaEq}).

\section{Perturbative expansion of $1/J$}\label{J}

The inverse of the determinant of the Jacobian matrix $\bJ$ can be expanded as  \begin{widetext}
\be
\frac{1}{J}=\text{Exp}[-\text{Log}(\text{det} (\mathds{1} + \bW))] = \text{Exp} \left[ -\text{Tr}\left( \sum\limits_{n=1}^{\infty} \frac{(-1)^{n+1}}{n} \bW^n\right) \right] = \sum\limits_{m=0}^\infty\frac{1}{m!}\left( \sum\limits_{q=1}^\infty \frac{(-1)^q}{q} \text{Tr}[\bW^q] \right)^m\,.
\ee 
This can be rewritten as
\be
\frac{1}{J}=\sum\limits_{m=0}^\infty \,\,\,\, \sum\limits_{k_1+...+k_\gamma=m} \,\,\, \prod\limits_{l=1}^\gamma \frac{(-1)^{l k_l}}{l^{k_l} k_l!} \left( \text{Tr}\left[ W^l \right] \right)^{k_l}\,,
\ee
\end{widetext}
where the second sum is taken over all the ordered partitions of $m$. It is not trivial to select a specific order of $1/J$ since we have to  substitute $W \rightarrow \sum_j W^{\text{(j)}}$, where $W^{\text{(j)}}$ is the order $j$ of the perturbed matrix $W$. Inserting this in thew above expression we obtain 
\begin{widetext}
\be
\left( \frac{1}{J}\right)^{(\al)}= \sum\limits_{m=0}^\al \,\,\,\, \sum\limits_{k_1+...+k_\gamma=m} \,\,\, \prod\limits_{l=1}^\gamma \frac{(-1)^{l k_l}}{l^{k_l} k_l!} \text{Tr}\left[    \sum\limits_{q_1+...+q_\ep=l} \,\,\, \prod\limits_{t=1}^\ep \frac{1}{q_t!} \left(W^{\text{(t)}} \right)^{q_t}   \right]^{k_l}\,,
\ee 
\end{widetext}
and now, for every ordered partition of $m$, select the terms that satisfy $\sum\limits_{l=1}^\gamma \left( \sum\limits_{t=1}^\ep t \, q_t \right) k_l = \al$ where for every $l$ the $q_t$ are an ordered partition of $l$. Up to third order this leads to the following expression for $J^{-1}$
\begin{widetext}
\be \begin{split}
\frac{1}{J} \simeq & \,\,1- \trr[W^{(1)}] +\frac{1}{2} \trr[W^{(1)}]^2 +\frac{1}{2} \trr[(W^{(1)})^2] - \trr[W^{(2)}] -\frac{1}{6} \trr[W^{(1)}]^3 -\frac{1}{3} \trr[(W^{(1)})^3] \\ &-\frac{1}{2} \trr[W^{(1)}] \trr [(W^{(1)})^2] +\trr[W^{(1)}W^{(2)}] + \trr[W^{(1)}] \trr[W^{(2)}] - \trr[W^{(3)}] + {\cal O}(4) \,.
\label{ea:det3ord}\end{split}
\ee
 \end{widetext}
\section{Standard results in LPT}\label{a:pert}

In standard LPT the Boltzman hierarchy is usually truncated setting $\sigma_{ij}=0$ so that the evolution of velocity dispersion is not considered and $S_\text{curl}=S_\text{div}=0$. The gradient and vector part of Euler equation in EdS can be written as the well known recursion relations (see main text for the definition of $\tau$) \begin{widetext}
\be
\begin{split}
 \left[ \frac{\pd^2}{\pd \tau^2} + \frac{1}{2} \frac{\pd}{\pd \tau} -\frac{3}{2} \right] \nabla \cdot \bpsi^{(n)} =& - \sum\limits_{\al+\b=n} \ep_{ijk}\ep_{ipq} \Psi_{j,p}^{(\al)} \left[ \frac{\pd^2}{\pd \tau^2} + \frac{1}{2} \frac{\pd}{\pd \tau} -\frac{3}{4} \right] \Psi_{k,q}^{(\b)}\\
& -\frac{1}{2}  \sum\limits_{\al+\b+\ga=n} \ep_{ijk}\ep_{pqr} \Psi_{i,p}^{(\al)} \Psi_{j,q}^{(\b)} \left[ \frac{\pd^2}{\pd \tau^2} + \frac{1}{2} \frac{\pd}{\pd \tau} -\frac{1}{2} \right] \Psi_{k,r}^{(\b)}\,,
\end{split}
\label{gradeq}
\ee
\be
 \left[ \frac{\pd^2}{\pd \tau^2} + \frac{1}{2} \frac{\pd}{\pd \tau} \right] \nabla \we \bpsi^{(n)} = \sum\limits_{\al+\b=n} \nabla \Psi_i^{(\al)} \we  \left[ \frac{\pd^2}{\pd \tau^2} + \frac{1}{2} \frac{\pd}{\pd \tau} \right] \nabla \Psi_i^{(\b)}\,.
 \label{curleq}
\ee  \end{widetext}

In the following we explicitly solve these equations up to second order with initial conditions for the displacement field to ensure that initial vorticity vanishes. We verify that at every order the vorticity is exactly vanishing, hence no vorticity is  generated. This result is not surprising if one looks at eq.~(\ref{vortexp}): if the Boltzman hierarchy is truncated in such a way that the velocity dispersion tensor is exactly zero, the source in eq.~(\ref{vortexp}) vanishes and the equation is homogeneous in $\omega$.  Therefore, if primordial vorticity  vanishes, in the absence of velocity dispersion, it remains zero throughout the entire cosmological evolution, as long as our fluid approach is valid.  

We list here the results for the leading order growing mode of the displacement field up to second order, while some details of the derivation are presented in the following sections.  In Fourier space 
  \be 
\tilde{\bpsi}^{(1)}(\bk) =i  \frac{\bk}{k^2} \,  \de_+ (\bk) D_+(\tau)\,,
 \ee
and 
 \be 
\tilde{\bpsi}^{(2)}(\bk) = i \frac{3}{14}\frac{\bk}{k^2} \, \alpha_{++} (\bk)  D_+(\tau)^2\,,
 \ee
where $\bk$ is the momentum associated to the Eulerian variable $\bx$ and we have defined 
\be
 \de_+(\bk)\equiv\frac{3}{5} \left(\de_0(\bk) -\frac{2}{3} \frac{\theta_0(\bk)}{ \HH_0} \right)\,,
\ee
\be
 \alpha_{++}(\bk) \equiv \int \frac{\dd^3 w}{(2 \pi)^3} \, \, \frac{(\bw \we \bk)^2}{w^2 |\bk-\bw|^2}  \de_+(\bw) \de_+(\bk-\bw)\,.
 \ee
In the matter dominated Universe which we consider here (EdS), the velocity divergence $\theta$ satisfies $\theta=-\mathcal{H}\delta$ and, given our normalization $a_0 = D_+(\eta_0) =1$, the quantity $\delta_+$ corresponds to the over density today, $\de_+\equiv \de_0$.

\subsection{  Initial conditions}\label{initial conditions}

One of the usual settings found in the literature for LPT initial conditions is the following: at first order one matches the initial value of the displacement field $\bpsi^{(1)}$ and its time derivative with the initial Eulerian fields and then one sets $\bpsi^{(n)}_{\text{in}}=0$ and $\dot{\bpsi}^{(n)}_{\text{in}}=0$ for $n \geq 2$. However this in return means that the initial conditions on the Eulerian fields at higher orders will in general  not be zero. In fact an Eulerian field at $n^{\text{th}}$-order at time $\eta_\text{in}$ will contain terms proportional to $(\bpsi^{(1)}\tin)^n$ that can not be compensated by higher order terms $\bpsi^{(n)}_{\text{in}}$ if we set them to zero.

To avoid this we will follow the first part of the usual description while we will use higher order initial conditions on the displacement fields to gain control on the Eulerian fields at $\eta\tin$ at every order. Motivated by the fact that inflation lays down small and linear perturbations we can treat the initial Eulerian fields as first order in LPT and set
\begin{align}\label{linear}
 \de\tin(\bx)=&\de\tin^{(1)}(\bq) \,,\quad \theta\tin(\bx)=  \theta^{(1)}\tin (\bq)\,, \\
 & \boldsymbol{\omega}\tin(\bx)=\boldsymbol{\omega}^{(1)}\tin(\bq)=0\,,
\end{align}
while at higher orders 
\be
\de\tin^{(n)} = \theta\tin^{(n)} = \boldsymbol{\omega}\tin^{(n)} =0 \quad \text{for} \quad n \geq 2\,.
\ee
In other terms, we chose to completely fix the initial Eulerian fields at every order and we impose initial conditions on the displacement field as a consequence of this choice. We have defined the initial (Eulerian) velocity divergence and vorticity as 
\begin{align}
&\theta_{\text{in}}(\bx) \equiv \left( \nabla_x \cdot \bv\right)_{\text{in}}\,, \\
& \boldsymbol{\omega}_{\text{in}}(\bx) \equiv  \left( \nabla_x \we \bv\right)_{\text{in}} \,.
\end{align}
In the linear equations (\ref{linear}), the Lagrangian and Eulerian variables are interchangeable: at linear order, the intrinsic non linearities of the Lagrangian approach are neglected and $\bx\sim \bq$.

The first order conditions $\de\tin^{(1)} = \de\tin$, $\theta^{(1)}\tin = \theta\tin$ and  $\boldsymbol{\omega}^{(1)}\tin=0$ can be implemented as conditions on the first order displacement\footnote{As we have already mentioned, the density field is not a dynamical variable in LPT. The continuity relation $\rho(\bx,\eta) \dd^3x= \bar \rho (\bq) \dd^3q$ implies $J (1+ \de)=1$, so that  
$\de\tin = 1/J\tin -1$.} (like in the main text, a prime denotes a derivative w.r.t  $\tau$)\small
\begin{align}
& \de	\tin^{(1)} = \left(\frac{1}{J\tin} \right)^{(1)} = - \nabla \cdot \bpsi\tin^{(1)} = \de\tin\\
&\theta_{\text{in}}^{(1)} = \left( \nabla_x \cdot \bv\right)_{\text{in}}=\left( \nabla_q \cdot \bv\right)_{\text{in}}  = \HH\tin (\nabla \cdot \bpsi^{(1)})_{\text{in}}' =\theta\tin\,,\label{initi}\\
& \boldsymbol{\omega}^{(1)}_{\text{in}} =  \left( \nabla_x \we \bv\right)_{\text{in}}= \left( \nabla_q \we \bv\right)_{\text{in}}  = \HH\tin (\nabla \we \bpsi^{(1)})_{\text{in}}'=0 \label{initi2} \,.
\end{align}\normalsize
 To implement the initial conditions on the higher order displacement fields we need expansions similar to the one we found for the vorticity in eq.~(\ref{vortexp}) for the $\de(\bx,t)$ and $\theta(\bx,t)$. For the density we can use eq.~(\ref{ea:det3ord}) and read out $\de^{(n)}$ up to the desired order while for the velocity field we obtain \begin{widetext}
\be\begin{split}
\theta^{(n)} =&\, a(t) \Biggl[ \sum\limits_{\al+\b = n} \left( \frac{1}{J}\right)^{(\al)} \dot \Psi_{p,p}^{(\b)} +  \sum\limits_{\al+\b+\gamma = n} \left( \frac{1}{J}\right)^{(\al)} \left( \Psi_{p,p}^{(\b)}\dot \Psi_{l,l}^{(\gamma)}- \Psi_{k,i}^{(\b)} \dot \Psi_{i,k}^{(\gamma)}  \right)\,\\
&+\frac{1}{2}  \sum\limits_{\al+\b+\gamma+\delta = n} \left( \frac{1}{J}\right)^{(\al)} \ep_{ijp} \ep_{kqr} \Psi_{j,q}^{(\b)} \Psi_{p,r}^{(\gamma)} \dot \Psi_{i,k}^{(\de)} \Biggr]\,.
\label{velexp}\end{split}\ee  \end{widetext}
The initial conditions are given at first order \be \begin{split}
& \nabla \cdot \bpsi\tin^{(1)} = -\de\tin\,,\\
& ( \nabla \cdot  \bpsi^{(1)} )'\tin =\HH^{-1}\tin \theta\tin\,, \\
& (\nabla \we \bpsi^{(1)} )'\tin = 0 \,,
\end{split}\label{firstIC}\ee
at second order
\be \begin{split}
& \nabla \cdot \bpsi\tin^{(2)} = \frac{1}{2} (\nabla \cdot \bpsi\tin^{(1)})^2 +\frac{1}{2} \Psi_{i,j}^{(1)} \Psi_{j,i}^{(1)} \Bigl|\tin\,,\\
& ( \nabla \cdot  \bpsi^{(2)} )'\tin = \Psi_{k,i}^{(1)} \Psi_{i,k}'^{(1)} \Big|\tin\,,\\
& (\nabla \we \bpsi^{(2)} )'\tin =\nabla \Psi_i^{(1)} \we \nabla_i \dot \bpsi^{(1)} \Big|\tin\,. 
\end{split}\label{secondIC}\ee

We point out that since $\boldsymbol{\omega}\tin$ is a pure vector the set of Eulerian initial conditions $\{\de\tin, \theta\tin, \boldsymbol{\omega}\tin \}$ has 4 degrees of freedom (at every order) while we are trying to constrain 6 degrees of freedom of the vectors $\bpsi\tin$ and $\bpsi'\tin$. This redundancy in the Lagrangian picture is already known and, as pointed out in  \cite{Matsubara:2015ipa}, the fact is that $(\nabla \we \bpsi)\tin$ has no effect on the Eulerian field: it is a time invariant rotation which simply relabels the initial coordinates. We are going to set, at every order, $(\nabla \we \bpsi)\tin=0$.\\

\subsection{First order}

At first order in the displacement field eqs.~(\ref{gradeq}) and (\ref{curleq}) reduce to
\begin{align}
&\left[ \frac{\pd^2}{\pd \tau^2} + \frac{1}{2} \frac{\pd}{\pd \tau} -\frac{3}{2} \right] \nabla \cdot \bpsi^{(1)} =0 \nn\\
&\,\,\,\,\, \rightarrow \,\,\,\,\,  \nabla \cdot \bpsi^{(1)} = c_1 e^{-\frac{3}{2} \tau} + c_2 e^{\tau}\,,
\end{align}
\begin{align}\label{c4}
&\left[ \frac{\pd^2}{\pd \tau^2} + \frac{1}{2} \frac{\pd}{\pd \tau} \right] \nabla \we \bpsi^{(1)} =0 \nn\\
&\,\,\,\,\, \rightarrow \,\,\,\,\,  \nabla \we \bpsi^{(1)} = -2 c_3 e^{-\frac{\tau}{2}} + c_4\,.
\end{align}
 Solving the coupled differential equations for the displacement field, with initial conditions given by (\ref{firstIC}) we find 
\begin{align}
 \nabla \cdot \bpsi^{(1)} =& -\frac{3}{5} \left(\delta_0 -\frac{2}{3} \frac{\theta_0}{\mathcal{H}_0} \right) D_+(\tau) -\nn\\
 &-\frac{2}{5} \left( \delta_0 +\frac{\theta_0}{\mathcal{H}_0} \right) D_-(\tau)\,,
\end{align}
\be
\nabla \we \bpsi^{(1)} = 
0 \,,
\ee\normalsize
where we have set to zero the constant mode $c_4$  in eq.~(\ref{c4}). This is a pure gauge mode which can be eliminated with a proper coordinate relabeling (see \cite{Matsubara:2015ipa}). We have defined (in EdS):
  \begin{align}
 &D_+ = e^\tau = a\,,\\
 & D_- = e^{-3/2 \tau}= a^{-3/2}\,,\\
 & E_-= e^{-\tau/2}=a^{-1/2}\,.
\end{align}
Using $\bpsi^{(n)}= \na^{-2} \left[ \nabla (\nabla \cdot \bpsi^{(n)}) - \nabla \we (\nabla \we \bpsi^{(n)}) \right]$ we can solve for $\bpsi^{(1)}$ directly in Fourier space and find~\cite{Matsubara:2015ipa}
\begin{align}
\ti\bpsi^{(1)} =& \frac{i}{k^2} \bk \left[ \frac{3}{5} \left(\delta_0 -\frac{2}{3} \frac{\theta_0}{\mathcal{H}_0} \right) D_+(\tau) +\right.\nn\\
&\quad \left.+\frac{2}{5} \left( \de_0 +\frac{\theta_0}{\mathcal{H}_0} \right) D_-(\tau) 
\right] \,,\label{firstpsik}
\end{align}
with $\theta_0 = i \, \bk \cdot  \bu_0(\bk)$, and $ \de_0= \de_0 (\bk)$. It is useful to rewrite $\ti \bpsi^{(1)}$ as
\be 
\ti \bpsi^{(1)} = \frac{i}{k^2} \left( \de_+ (\bk) D_+(\tau)+  \de_- (\bk)  D_-(\tau) \right) \bk \,, \label{psi1DD}
 \ee
 with 
 \begin{align}
  & \de_+(\bk)\equiv\frac{3}{5} \left(\de_0 -\frac{2}{3} \frac{\theta_0}{\HH_0} \right)\,,\\
  &\de_-(\bk)\equiv\frac{2}{5} \left(\de_0 +\frac{\theta_0}{ \mathcal{H}_0} \right)\,.
 \end{align}
 The result in real space is given by 
 \be 
\bpsi^{(1)} = -\frac{\nabla}{\Delta} \left( \delta_+ (\bq) D_+(\tau)+ \de_- (\bq) D_-(\tau) \right) \,,\label{psi1real}
 \ee
 where $\Delta^{-1} $ denotes the inverse Laplacian.
In the first order expression above we can treat $\bq\sim \bx$ and consider the $\bq$-Fourier space as the Fourier space associated to Eulerian coordinates, for the reasons explained in section \ref{initial conditions}. 

We point out that if we were to impose $\omega\tin \neq 0$ we would have an additional mode in the first order displacement field proportional to the initial vorticity and with time dependence $E_-(\tau)$. We can quickly compute the first order vorticity
\begin{align}
&\omega_\ell^{(1)} = \, a(\eta) \left( \frac{1}{J}\right)^{(0)}  \ep_{\ell kj}\dot \Psi_{j,k}^{(1)} \nn\\
& \,\,\,\,\, \Rightarrow \,\,\,\,\, \boldsymbol{\omega}^{(1)} (\bk, \eta)=\HH \left( \nabla \we \bpsi \right)' = 
0\,,
\end{align}
 and we see that there is no generation of vorticity at first order.\\

\subsection{Second order}

We now move to the second order solutions. The recursion relations at second order yield
\begin{align}
&\left[ \frac{\pd^2}{\pd \tau^2} + \frac{1}{2} \frac{\pd}{\pd \tau} -\frac{3}{2} \right] \nabla \cdot \bpsi^{(2)}=\nn\\
& =-\ep_{ijk}\ep_{ipq} \Psi_{j,p}^{(1)} \left[ \frac{\pd^2}{\pd \tau^2} + \frac{1}{2} \frac{\pd}{\pd \tau} -\frac{3}{4} \right] \Psi_{k,q}^{(1)}\,,\label{recurgrad2}
\end{align}
\begin{align}
&\left[ \frac{\pd^2}{\pd \tau^2} + \frac{1}{2} \frac{\pd}{\pd \tau} \right] \nabla \we \bpsi^{(2)} =\nn\\
&= \nabla \Psi_i^{(1)} \we  \left[ \frac{\pd^2}{\pd \tau^2} + \frac{1}{2} \frac{\pd}{\pd \tau} \right] \nabla \Psi_i^{(1)}\,.
\label{recurtran2}
\end{align}
Enforcing initial conditions (\ref{secondIC}), we obtain
 \begin{widetext}
\be
\begin{split}
\bk \cdot \ti \bpsi^{(2)} =& 3  i \,\alpha_{++}(\bk) \left( \frac{e^{-\frac{3}{2}\tau}}{35}-\frac{e^\tau}{10}+ \frac{e^{2\tau}}{14} \right)+i \alpha_{--}(\bk) \left(\frac{e^{-3\tau}}{8} -\frac{e^{-\frac{3}{2}\tau}}{5}+\frac{3e^\tau}{40} \right)+i \, \alpha_{+-}(\bk) \left(\frac{3e^{-3\tau/2}}{5} -e^{-\tau/2}+\frac{2e^\tau}{5} \right)+\\ 
&+ \frac{2}{5} \left[ (\bk \cdot \ti \bpsi^{(2)})\tin -(\bk \cdot \ti \bpsi^{(2)})'\tin \right] e^{-3\tau/2}  +\frac{1}{5} \left[ 3(\bk \cdot \ti \bpsi^{(2)})\tin +2(\bk \cdot \ti \bpsi^{(2)})'\tin \right] e^{\tau} \,,
\end{split}\ee
\be
\bk \we \ti \bpsi^{(2)} = 2 \left(1-e^{-\tau/2} \right) (\bk \we \ti \bpsi^{(2)})'\tin= 5 i  \left(1-e^{-\tau/2} \right) \boldsymbol{\beta}_{+-} \,,
\label{curlmode2}
\ee
\end{widetext}
 with
 \small
\be
\begin{split}
& \al_{AB}(\bk) \equiv \int \frac{\dd^3 w}{(2 \pi)^3} \, \, \frac{(\bw \we \bk)^2}{w^2 |\bk-\bw|^2}  \de_A(\bw) \de_B(\bk-\bw) \,, \\
&\boldsymbol{\b}_{AB}(\bk) \equiv \int \frac{\dd^3 w}{(2 \pi)^3} \, \, \frac{(\bw \we \bk)}{w^2 |\bk-\bw|^2}  \bw \cdot (\bk-\bw) \, \de_A(\bw) \de_B(\bk-\bw) \,.
\end{split}\label{dd2}
\ee 
\normalsize
The indices here take the symbolic values $A,B=(+,-)$ and we point out that $\al_{AB}$ and $\boldsymbol{\b}_{AB}$ are respectively symmetric and antisymmetric in the exchange of the indices.\footnote{We have already used $\boldsymbol{\b}_{++}=\boldsymbol{\b}_{--}=0$ and $\boldsymbol{\b}_{+-}=-\boldsymbol{\b}_{-+}$ for the simplification that led to eq.~(\ref{curlmode2}). We stress that if we assume that the density perturbations are independent of direction, i.e. $\de_i(\bk)=\de_i(|\bk|)$, we have $\boldsymbol{\b}_{ij}(\bk) =0$. This can be understood as follows: in this case the only vector that can appear in the $\boldsymbol{\b}_{ij}(\bk)$ is $\bk$ so that we have $\boldsymbol{\b}_{ij}(\bk) \propto \bk$ but since this is supposed to be a curl-mode, see eq.~(\ref{dd2}), they must be zero. This is expected if we assume isotropy of density perturbations.} We underline that in eq.~(\ref{dd2}) the Fourier space can be considered as the Fourier space associated to the Eulerian coordinates $\bx$: being $\de_A(\bq)$ a first order quantity we can treat it as a function of $\bx$.

We can now use eq.~(\ref{vortexp}) to write 
\be
\omega_\ell^{(2)}= \HH(\eta) \left(  (\nabla \we \bpsi^{(2)})'_\ell + \ep_{\ell kj} \Psi_{i,k}^{(1)}  \Psi'^{(1)}_{j,i} \right)\,,
\label{secondvortzero}
\ee
which upon substitution of the the first and second order solutions for the displacement fields gives
\be\
\boldsymbol{ \omega}^{(2)} (\bk, \eta)= 0  \,,
\label{vort2}
\ee
there is no generation of vorticity also at second order.  

\subsection{Higher orders}

One can in principle continue this iterative procedure to compute $\boldsymbol{\omega}^{(n)}$ at every order. However, if proper initial conditions are enforced on the displacements fields such that $\boldsymbol{ \omega}^{(n)}\tin=0$, eq.~(\ref{vortexp}) has to be solved without a source and  it yields $\boldsymbol{\omega}^{(n)}(\bx, \eta)=0$ and any order $n$.

\newpage
\bibliographystyle{utcaps}
\bibliography{vorticity-refs}

\end{document}